# ROLE AND DISCIPLINE RELATIONSHIPS IN A TRANSDISCIPLINARY BIOMEDICAL TEAM: STRUCTURATION, VALUES OVERRIDE AND CONTEXT SCAFFOLDING

Gaetano R. Lotrecchiano, EdD, PhD

George Washington University, Department of Clinical Research and Leadership
School of Medicine and Health Sciences
Washington, DC, 20037
e-mail: glotrecc@gwu.edu

**ABSTRACT**

Though accepted that "team science is needed to tackle and conquer the health problems that are plaguing our society"[1] significant empirical evidence of team mechanisms and functional dynamics is still lacking in abundance. Through grounded methods[2] the relationship between scientific disciplines and team roles was observed in a United States National Institutes of Health-funded (NIH) research consortium. Interviews and the Organizational Culture Assessment Instrument (OCAI)[3] were employed.. Findings show strong role and discipline idiosyncrasies that when viewed separately provide different insights into team functioning and change receptivity. When considered simultaneously, value-latent characteristics emerged showing self-perceived contributions to the team. This micro/meso analysis suggests that individual participation in team level interactions can inform the structuration of roles and disciplines in an attempt to tackle macro level problems.

**INTRODUCTION**

Academic disciplines have traditionally operated as guilds with the power to confer or discourage pedigree upon its members. As such they recognize and manifest an individual's identity controlled by constructs of any given time, place, or group embedded with values sourced in the culture of the discipline or profession. The actual acquisition of disciplinary knowledge is often subordinate to the socializing elements that organize and redistribute it.

Conducting transdisciplinary (TD) science in biomedical healthcare teams has become increasingly more important as scientists and stakeholders strive for increased collaborations between and across disciplinary silos.[4] "There is ample evidence that when scientists work cooperatively with knowledgeable activists from outside the research community, science as well as society can benefit".[5] However, biomedical science has historically been characterized within compartmentalized units of inquiry often within closed systems of knowledge resisting epistemological and methodological change grounded in unidisciplinary value systems.[6] Specifically, basic-applied sciences have been hampered by a unidirectionality, providing solutions to problems from a position of control and the isolation of the laboratory.[7] This has caused a lack of interest in interacting with adjacent communities and resistance to change that may lead to the successful generation of new knowledge for the healthcare enterprise and value inculturation. The growth of inter- and TD communities working within pluralistic interdisciplines has been stalled as a result.[1] We therefore ask several questions of TD science teams. First, what is the role of values within the diversity of TD biomedical team? Second, what are the change relationships between the micro and meso levels of this sort of enterprise? And third, what are the mechanisms that build context for teams.

**LITERATURE REVIEW**

**The emerging collaborative landscape.** U.S. imperatives for TD biomedical research is a phenomenon encouraged by the NIH's now decade-long focus on translational science, the bold approach that challenges traditional biomedical scientific social structures[8] and encourages the science-of-team-science (SciTS).[9,10] These co-developing preoccupations are grounded in innovation and integration of efforts. Team science expresses itself through mechanisms that ensure outcomes of transformed, novel, and complex structures for complex problem solving. In healthcare, patients, politicians, advocates, and other stakeholders usually absent from the biomedical research agenda are now invited into conversations about scientific matters to add credibility, relevance, and to increase significance. The slowly growing collection of evidentiary research on how biomedical teams operate and are managed provides social scientists with the tools needed to utilize historic and emerging theories that further research team enterprises beyond their natural barriers and generates viable solutions to difficult to solve problems.

**Team member familiarity and social cohesiveness.** Issues related to team cohesion are found in both the social psychology and management literature and attempts to measure effectiveness and decision-making.[11,12] Some researchers have concluded that cohesion is directly related to heightened performance and that "good" or desirous performance reciprocates cohesion. Contrarily, other cases found that homogeneity in groups, though more likely to breed cohesion, is less likely to be more successful in some tasks as heterogeneous groups.[13-15] It has been found that there is a longitudinal vector associated with these findings suggesting that familiarity of team members over time may affect the desire to maintain high levels of performance. [12,16] Some studies have found that this is partially due to emergent social behaviors that are bred through familiarity like social loafing, and groupthink[17,18] that may be deterrents to high performance. Familiarity may in fact breed inflexibility over time as communication skills decline.[16] Some continue to argue that this is due to the reluctance to modify pre-established roles and interaction patterns[19] transversely, attempt to survive changing environments as team members evaluate outdated behaviors and abandoning them for new ones when needed. [20]

**Role/discipline relationships, values, and culture.** The problem of mapping role/discipline relationships comes from the hazard of associating one's role with their disciplinary affiliation.[21] This relationship has shown to be interlinked with team cultures and values.[22-24] It is often assumed that a group or team works from one cultural framework rather than multi-valued backgrounds resulting from multiple disciplinary traditions in conversation. There may be a proportional relationship between conflict, diversity, and team outputs[25,26] for positive innovative results.

The consideration of values as one type of mechanism within teams helps to understand how diverse values and motivations may affect functionality when teams are faced with shared goals and outcomes. Technical, practical-hermeneutic, and emancipatory interests serve as logical motives for seekers of knowledge in a team project. Technical interest is most suited to those hoping to gain some improved performance from the study of organizational culture. "Advocates of this view believe that it is vital to uncover linkages or causal relationships between forms of organizational culture and performance".[24] Process in collaboration is paramount in this interest as team members focus on *how to interact* most effectively. The practical-hermeneutic interest is more prone to include interests in the meaning of certain organizational communities. Causality is secondary in this interest and the paradigm is more focused on the *meaning of interacting* as a group. Knowledge is in itself a useful outcome but not paramount. Finally, emancipatory interest focuses on the "so-called" negative features of organizational culture in the hopes of rectifying it to generate more appropriate values systems within an organizational culture.[27] Freedom from repression and enacting free will are operative in this motif. Team members assume that interactions expand the scope of knowledge by inclusion of an expanded stakeholder community. Each of these is useful in the organization and understanding of meaning for certain types values that may be operative within team interactions.

Mixed motivations and shared outcomes prove to provide a level of conflict that is both multi-leveled and contributes differently to structuration of the team. These mixed motivations are linked to a diversity of values and the cultures (or disciplines) that they represent.[22-24] They are not only proportional to team outputs[25,26] but are emergent and interactive by nature[28] thus collectively contributing to future mesostructures as micro level tensions between actors and their own discipline, and actors and the collective, continually generate tension.[29,30] Questions concerning the nature of transdisciplinarity and the development of inter-disciplines that are a result of it, those that represent more than one historical, methodological, or epistemological tradition (epigeneticists, pediatric neurologists, neuroanatomists, etc.) are only one consideration. Boundary crossing and the ability to ensure constant evolutionary states of knowledge emergence as process and outcome is another. And the role of collaboration and coordination in such teaming environments is even yet another.

**Structuring interactions and desired change within teams.** Change relationships between micro (individual) and meso (team) levels requires focus on the structuring interactions that occur between levels[30] Structuration theory has a dualistic view of individual roles within systems that "not only monitor continuously the flow of their activities…they also routinely monitor aspects of the contexts in which they move".[31] Individuals draw on both the mundane everyday attributes of their discipline and role oriented constructs while simultaneously informing them as part of the same structuring process.[32] This process allows for micro level exchanges between individuals and their respective disciplines to become part of the change process within knowledge communities.[33]

Collaboration is more illusive when role/discipline relationships are considered within TD teams. Studying interactions within teams of stakeholders is a qualitative inquiry that searches for the identification and meaning of certain social mechanisms.[34] This is relevant because TD is an unlikely constant state in social systems. It may at times be a novel outcome of team science collaborations and at other times may be dormant to the more apparent multi- and interdisciplinary interactions imbedded within group.[35]

Buckley's approach and model are helpful within this interplay between reference groups. Cultural material transitioning from actor to actor is less important than the "triggering" effects of information transmission that are generated by them. Buckley[36,37] focus on the socio-cultural systems as "carriers of meaning rather than entities that exist some place or flow from one place to another".[38] This morphogenic view is less structural and more interactive and the measurable units of activity are relational and dynamic rather than codifiable and static. It introduces the possibility of theory based on conversation and discourse. "Conversation theory explicitly builds a bridge between studies of social systems and social behavior and studies that investigate the interactions of social actors to approaches that attempt to characterize social systems as distinct forms of autonomous wholes".[39] It is an attempt to introduce modern systems theory to social science underscoring the ultimate impact conforming and non-conforming individuals have on the society as a whole.

## METHODS

The population of the NIH scientific consortium studied was 70 representing 14 centers located throughout the US and abroad. Participant selection was based on ensuring representation of the multiple centers, a parent/patient organization, and the stratification of disciplines and social roles from within the group. Table 1 describes the population, the stratifications, and proposed prioritization criteria. Table 2 describes the two phases of data collection and the sample obtained.

**Phase 1: Modified Competing Values Framework.** The Organizational Culture Assessment Instrument (OCAI) is a meta-theoretical construct designed to measure in-group values as they contribute to team construction. This framework allowed the examination of cognitive structures, intra-organizational tensions caused by conflicts between beliefs, and values found in the group. Through a modification of this tool, a view of the team culture was captured and mapped. The tool assisted in understanding the role individual values play in the team and how these values described an overall interpretation of the team.

Table 1. Population, Stratification, and Prioritization of Participants

| Population | Sample Stratification 1 (Discipline) | Sample Stratification 2 (Social role) | Prioritization Criteria |
|---|---|---|---|
| National Consortium w/international partners (14 centers) (1 parent organization) | Management/Finance (A) Neuro-Behavioral (B) Metobal-Nutrition ( C) Parent-Pediatrics (D) Genetics (E) | Administration (1) Advocate (2) Principal Investigator (3) Clinical Associate (4) Researcher (5) | - Stratification 1 - Stratification 2 - Diverse center representation |

Table 2. Sample Population

| Collection Source | Sample |
|---|---|
| **SURVEY:** Organizational Culture Assessment Instrument (OCAI). | N=20 |
| **INTERVIEWS:** Validation of survey data. Description of meanings. | N=10 |

Figure 1 illustrates the OCAI Framework.[40] Two competing dimensions are featured as X and Y axes. The X-axis represents the tension between the spontaneity and flexibility. The Y-axis represents the tension between internal demands for system maintenance and external productivity and competition-driven demands.

Through mapping the interplay between each of these demands and the tensions they create, four primary cultural quadrants are highlighted: the human relations or clan model (C) representing a group-oriented culture that values humanism (upper left), the open systems or adhocracy model (A) representing a more transforming and developmental culture focused on growth (upper right), the hierarchical model (H) seen in cultures that value balance, control and continuity (lower left), and the rational goal or market model (M) representing an output-focused, goal-driven culture that values accomplishment and efficiency (lower right).

Zones emerge generating a visual tool that discloses the underlying values and assumptions in a given group culture now and in a preferred state. "The result is a holistic profile of organizational culture".[41]

**Phase 2: Participant interviews.**
Interviews were conducted in person, via telephone, or through the online posting mechanism with individuals of the team. Purposeful and snowball sampling[42] was employed to match diversity in disciplines and roles. The design of the interview protocol was semi-structured (Table 3 shows the main questions), allowing for focus on the phenomena itself and the emergent details of description.[43] Interviews were coded thematically and organized through the ATLAS.ti software package.

Triangulation occurred by cross-referencing reference groups (termed clusters) and participant data with semi-structured interview answers. Participant interviews played a large role in checking the researcher's interpretation of focus group and survey material. Data was subject to external audits by colleagues possessing adequate skill to assess the effectiveness of the design and appropriateness of analysis. Internal auditing was a fixed strategy within the participant group as data and analysis was continually audited and member checked by participants and in individual interviews. All data was de-identified.

Figure 1. Competing Values Framework.[40]

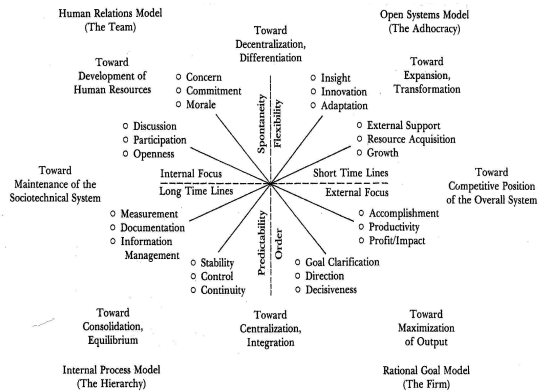

Table 3. Individual Interview Questions.

| |
|---|
| What constitutes TD in team science? |
| How is successful team science achieved (or not)? |
| Do values play a supporting or disarming role building TD team science? |
| Do values play a supporting or disarming role building TD team science? |
| What are the social mechanisms? |
| How do social mechanisms emerge? |

## RESULTS

In the analysis, emphasis has been placed on disciplinary and role based differences and similarities between cluster groups of disciplines and roles concerning 1) degree of culture change preference (change factor), 2) preferred culture, 3) the function of the individual, 4) the individual relationship to the team, and 5) discipline/role conflict. Table 4 outlines the results that these data once analyzed provides. The relationship between role and discipline, the cultural preferences of the participants, and change factors show how the patterns when considered simultaneously yield information about individual participant's role and discipline in light of the entire team make up. This is depicted in the radial graph in Figure 2. Identification of each participant's role and discipline, now and preferred culture, the direction of culture change and also if he or she is part of the no, low, moderate or high change factor within the group is shown.

Table 4. Key evidence from data (micro level)

| Key Evidence from Data (Micro Level) *displayed in order of change degree of change preference (high to low)* | | | | |
|---|---|---|---|---|
| *Discipline Cluster Group* | *Degree of change* | *Preferred culture* | *Function of the individual* | *Individual's relationship with the group* |
| **Neuro-Behavioral (NB)** | High | Internal focus and Integration | Part of the collective | Contributing to the collective |
| **Parent-Pediatrics (PP)** | Moderate | Discretion and External Focus | (Role/Discipline Overlap) | Manage expectations for the group |
| **Genetics (GE)** | Moderate (with high diversity) | Flexibility, External Focus and Differentiation | Affected by high trust expectations | Strong clan dependency |
| **Management/Finance (MF)** | Low to Moderate | Stability and control | Affected by hierarchy | Provides stability |
| **Metabol-Nutrition (MN)** | Low | Internal focus and Integration | Affected by hierarchy (Role/Discipline Overlap) | Assignment/Service orientation to the group |

| Key Evidence from Data (Micro Level) *displayed in order of degree of discipline/role conflict (high to low)* | | | | |
|---|---|---|---|---|
| *Role Cluster Group* | *Degree of change* | *Preferred culture* | *Function of the individual* | *Disc/ Role Conflict* |
| **Advocate (AV)** | Moderate | External focus and differentiation | Suppression of discipline in favor of role. | High |
| **Administrative (AD)** | Moderate | Flexibility, discretion and internal focus. | Exact role-driven (Discipline/Role overlaps) | Low |
| **Principal Investigator (PI)** | None to High | Flexibility | Relationship building (Relationships affect success) | Low |
| **Clinical Associate (CA)** | Moderate | Flexibility, External focus, differentiation, & discretion | Affected by hierarchy | Low |
| **Researcher (RE)** | Low to moderate | Flexibility and integration | Discipline/Role divisions (Ext. factor dependent) | Low |

Figure 2. Role/Discipline Culture and Change Factors

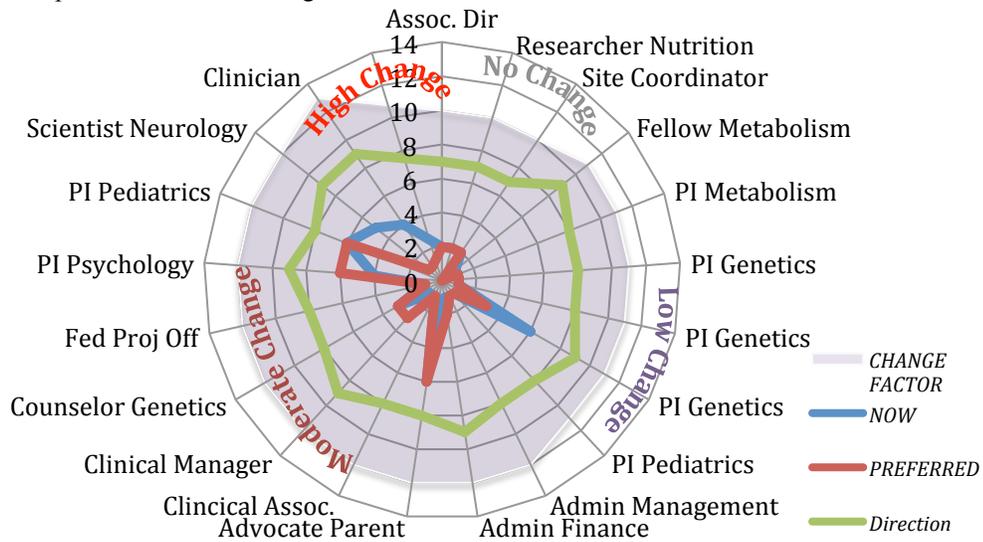

*Legend*

    *Culture quadrants:* Balanced=0, Clan/Adhocracy=1, Clan/Hierarchy=2, Clan/Market=3, Hierarchy/Adhocracy=4, Hierarchy/Market=5, Market/Adhocracy=6.
    *Directional Change:* No directional change (equal)=7, reinforcement of culture=8, towards another culture=9
    *Change Factor:* No change=10, Low change=11, Moderate change=12, High change=13

Figure 3. Conceptual Framework

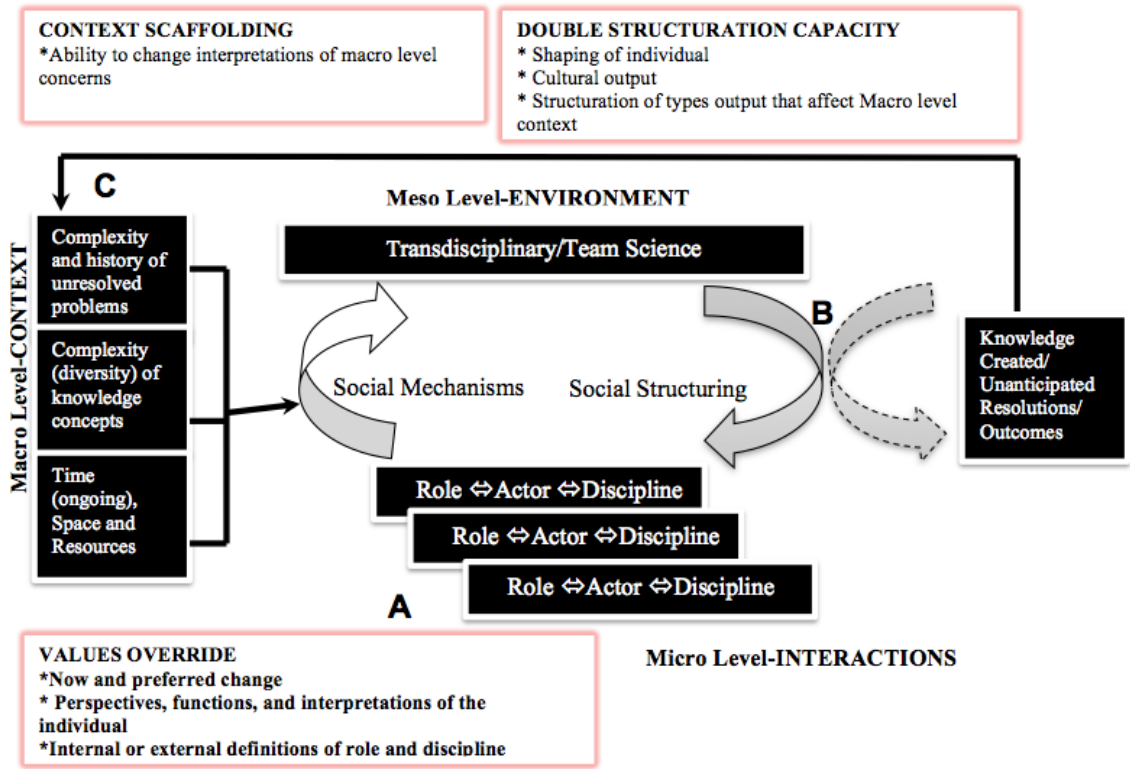

## DISCUSSION

*Proposition 1: Shared values override structural barriers in teams.*

**Values Override.** The usefulness of the OCAI has been that it allowed the research to map value preferences in a very easy and accessible way for this sample of individuals. Part of what was learned from utilizing this tool along with the other techniques of was that when considered separately, the differences and similarities between role and discipline perceptions of team involvement are visible, allowing for evidence of diverse value systems.[22-24] It provided a glimpse into how different clusters viewed the culture but also the preferred culture and what this might say about their cluster group. When disciplines/roles perspectives were overlapped, participants' interviews indicated much deeper descriptions of a values overriding, where their team values provided overarching reasoning beyond their disciplinary or role perspectives.

Typically, team performance is focused on task and goal management. The team science characteristics shown in this research measures team members' effectiveness in a team based on their mastery over the essential components of being a team "member" (as a metaphorical part of the body) and not just being part of the team (as a replaceable mechanical element) providing purpose based values. This sort of mastery is an outcome of learning[44] that is focused on mastering team identity and not only task fulfillment. It is a systemic consideration that has macro level implications as the individuals learns such values from team involvement and thus returns such mastery to the team providing the potential of a changing macro level perceptions of purpose (Figure 3, A.).[45,46]

**The Value of the "Values Override".** The values override, the result of role/discipline overlap, provides and important glimpse into the perceptions of individuals that are divorced yet intimately a part of their own role and/or disciplinary system of thought and team organization. It is a truly micro-meso level intersection of knowledge. Descriptions of perceptions on this level have the potential to be rich explanations of why, how, and with what vision individual team members see the own place within a working group. It is a part of the socio-cultural system that Buckley refers to as "carriers of meaning rather than…some place or flow from one place to another".[38] When accessed, the values override assist researchers in understanding the role of individual perceptions on a new level and define the congruence and divergence with team level values. Furthermore, the gathering of this sort of data, allows for a better understanding of team values that is unilateral and defined by leaders or long-time founding members of the group. This is important if one is to expect that change and value negotiation will become constant aspects of team engagement.

*Proposition 2: Team involvement has a direct effect on individual self-perception of role and disciplinary impact and can inform macro level problem solving.*

**Output occurrences of structuration.** Output occurrences from the team (meso level) affect the individual (micro level). Much has been learned from the mapping of preferred culture and how these infer the impact of team structures on individual's self-perceptions. Individual influence on the team varies and has been measured in the form of individual perceptions of role. What is visible is that the team plays a large part in shaping the functioning of the individual.[31] As such, what is not visible is how this influence changes individual self-awareness. This relationship requires further research. The structuring mechanism at play between the meso and micro levels can be observed when the preferred culture is measured as an indicator of what change the individual needs from this influencing agent.[38] Thus for instance, if a preferred culture indicates a movement away from clan/adhocracy (C/A) to market/adhocracy (M/A), it is possible to deduce that an individual may require change on the part of the team toward a market outlook therefore denoting that the social structuring mechanism operative is one that is negatively received by the individual. It is also possible that if an individual member shows a preferred desire to reinforce a C/A culture, the social structuring mechanism is one that will exponentially reinforce the team as is mediated by the individual through the social mechanisms operative in the team as input.[31]

**Double Structuration Capacity.** Such feedback from the meso level to the micro level as described above is part of the structuration of individuals and their identity. This confirms Giddens' assertion against the effectiveness of micro/macrosociology due to its inablity to arrive at conclusions such as those that can only be observed in the micro-meso level bridge. What has been found through this research is that individual participation in teams is transformational in that it goes beyond "one's taken-for-granted mode of being…subverted or unsettled, mak[ing] one suddenly conscious of that which was previously pre-reflected".[47] In this context, the values override is a testament to this finding.

In addition to the effect of structuration on individual team member mastery, we can also conjecture that this process is critical to the altering of outcomes from teams who affect macro level agendas.[25,26] This dense awareness of another reality besides one's own disciplinary and role oriented contributions, *de facto*, altars future outcomes[31,47] that can shape macro level perceptions of problems, complexity, and time and space considerations. This awareness does not just have theoretical implications but practical ones as well.

A double structuration capacity is therefore the ability to affect change on both the individual and outcome levels. Individual change is instilled as a product of mastery over team values that become individual ones. The byproduct of this transformative effect can over time effect macro level structures (Figure 3, B.).

*Proposition 3: Individual/Team context is a scaffolding experience of change that affects macro level perceptions of complex problems.*

**Context Scaffolding.** We have learned what the process of a micro/mesosociology contributes to a micro/macro one. Double structuration leads us to a conclusion that Giddens was correct that micro/macrosociology is unable to generate certain context specific realities that inform how transformation can affect interactive enterprises. Though this is the case here, we consider scaffolding of new perceptions and considerations of macro level concerns through the transformation of individuals and outcomes from teams.[25,26] These individual and outcome changes are "system-wide actions".[28] The micro- and meso-levels become part of the scaffolding which provides for "rich fields of emergent knowledge"[48] as system level processes are routed from within by individual and team oriented clusters (Figure 3, C.).

We can see how individual mastery[44] over team outputs that reshape individuals can affect macro level concerns. As individuals are transformed through their reciprocal relations[47] with others in teams they in turn affect team values and outcomes. As these outcomes affect macro level concerns they bring with them the individual reality shifts, methodological pluralisms, collaborative deconstructions, praxis, and interpenetration of epistemologies presented earlier that have the power to reorient the meaning of macro level questions and transform them. This is grounded in shifting motiviations that occur during the process of transformation as one moves from being a dynamic individual to a team member back to an individual. This cycle has the power to reshape motivations and inform macro level problem solving as a resultant output.[25,26],[31]

**CONCLUSION**

There are several important implications for further research and the identification of deeper more value-rich considerations in team dynamics.

First, role-driven and discipline-driven identities in teams provide only limited and possibly contradictory material when considering the functioning of team members. Such isolated considerations can easily divert one's attention away from the combination of characteristics and dynamics that are at play in a TD team based on *both* discipline-specific and role-specific material. This study has provided a preliminary basis for further research that focuses on these overlaps and their meaning. Deemphasizing role and discipline distinctions are key to more holistic approaches toward teams.

Second, the overlap of role and discipline dynamics provides an important opportunity to study less researched areas like values in teams that seem to override the discipline and role dichotomies. This could very well be the next important transition for team science studies that have historically moved from social structuring dynamics, to social psychology and organization science,[10] to studies which are more so grounded in a focus on micro-meso interactions of individual level contributions to meso level meanings and the role of this exchange on structuration of teams.[47]

Thirdly, an important consideration in TD teams is role expansion. Role release involves assigning specific tasks held by one team member over another.[49] Role expansion on the other hand is more like what is suggested in TD teams. It emphasizes goals of teaming and cross training and fits with the structuration aspect of this study. Here individual roles and disciplines are subject to and evolution that through certain mechanism will alter individual's knowledge, skills and abilities over time as a result of this type of teaming. This study promotes future team science constructs and identity development especially with regard to growing membership, sharing influence, developing leadership, and nurturing new personnel within teams. These identity structures need not only be relegated to the role and discipline traditions that shape team member's awareness of the teaming enterprise and their potential contributions to it, but could be useful in understanding better how their own personal characteristics might add to such dynamics within teams like conflict resolution, inequities, power imbalances between members, and the like.

Fourthly, the measures offered here are useful for certain teams if measurement of dynamics becomes more commonplace. Instead of creating evaluation processes that measure effectiveness through outcome orientations, teams can be measured through more topographic means that emphasize the peaks and valleys (triggers) of effectiveness based on dynamic interventions.[28] Effectiveness in this case could have multiple data points ranging from standardized goals, team specific ones, or even interpersonal ones. Longitudinal measures could also be employed that monitor team activity with a cybernetic lens capturing the relational coordinates of individuals, roles and disciplines, and episodes of breakthrough and progress and the shifting of strengths and weaknesses over periods of evaluation. Though speculative, these types of measures are not unavailable.

Lastly, the distinction between the meaning of discipline-specific and role-specific characteristics that this study showed provides evidence that much more of the current research into team science needs to incorporate multi-level evidence so as to inform the important questions about how team collaborate.[50] This stated, trends are shifting rapidly. The inequality and imbalance of team membership shown in this study are only beginning to be incorporated into this literature since such dynamics are usually applied to the description of intra-organizational and inter-sectoral partnerships.

**REFERENCES**
1. Disis, M. L. & Slattery, J. T. The road we must take: Multidisciplinary team science. *Science Translational Medicine* **2**, 22-29 (2010).
2. Borner, K. *et al.* A multi-level stystems perspective for the science or team science. . *Science Translational Medicine* **2**, 1-5 (2010).
3. Cameron, K. S. & Quinn, R. E. *Diagnosing and Changing Organizational Culture*. (Jossey-Bass, 2006).
4. Wuchty, S., Jones, B. F. & Uzzi, B. The increasing dominance of teams in production of knowledge. *Science* **316**, 1036-1039 (2007).
5. Sarewitz, D. Social change and science policy. *Issues in Science and Technology* **13**, 29-32 (1997).
6. Berisio, R., Baram, D. & Keinan, E. Nobel poze in chemistry 2009: When biology turns into chemistry. *Isreali Journal of Chemistry* **50**, 24-26 (2010).
7. Fiore, S. Power and promise: Cognitive psychology and cognitive technology. *Cognitive Technology* **13**, 5-8 (2008).
8. Zerhouni, E. The NIH Roadmap. *Science* **301**, 63-72 (2003).
9. Stokols, D., Hall, K., Taylor, B. K. & Moser, R. The science of team science. *American Journal of Preventative Medicine* **35**, S77-S88 (2008).
10. Stokols, D., Misra, S., Moser, R., Hall, K. & Taylor, B. K. The ecology of team science. *American Journal of Preventative Medicine* **35**, S96-S115 (2008).
11. Kerr, N. L. & Tindale, R. S. Group performance and decision making. *Annual Review of Psychology* **55**, 623-655 (2004).
12. Guzzo, R. A. & Dickson, M. W. Teams in organizations: Recent research on performance and effectiveness. *Annual Review of Psychology* **47**, 307-333 (1996).
13. Milliken, F. J. & Martins, L. L. Searching for common threads; understanding the multiple effects of diversity on organizational groups. *Academy of Management Review* **21**, 192-199 (1996).
14. Wiersema, M. F. & Bantel, K. A. Top management team demography and corporate strategic change. *Academy of Management Journal* **35**, 91-121 (1992).
15. Jackson, S. E., May, K. E. & Whitney, K. in *Team Effectiveness and Decision Making in Organizations* (eds R.A. Guzzo & E. Sales) 204-261 (Jossey-Bass, 1995).
16. Katz, D. The effects of group longevity on project communication and performance. *Administrative Science Quarterly* **27**, 81-104 (1982).
17. Janis, I. *Groupthink: Psychological Studies of Policy Decisions and Fiascoes*. (Houghton Mifflin, 1982).
18. Kayes, A. B., Kayes, D. C. & Kolb, D. A. Experiential learning in teams. *Simulation Gaming* **36**, 330-354 (2005).
19. Okhuysen, G. A. Structuring change: familiarity and formal interventions in problem-solving groups. *Academy of Management Journal* **44**, 794-808 (2001).
20. Weick, K. E. *The Social Psychology of Organizing*. (Addison-Wesley Publishing Co. , 1979).
21. Ashforth, B. E., Rogers, K. M. & Corley, K. G. Identity in organizations: Exploring cross level dynamics. *Organizational Science* **22**, 1144-1156 (2011).
22. McFeat, T. *Small-Group Cultures*. (Pergamon Press, 1974).
23. Sackman, S. A. Culture and subcultures: An analysis of organizational knowledge. *Administrative Science Quarterly* **3**, 140-161 (1992).
24. Alvesson, M. *Understanding Organizational Culture*. (SAGE, 2002).
25. Jehn, K. & Chatman, J. A. The influence of proportional and perceptual conflict

26 Jehn, K., Northcraft, G. B. & Neale, M. A. Why differences make a difference: A field study of diversity, conflict and performance. *Administrative Science Quarterly* **44**, 741-763 (1999).

27 Habermas, J. (Heinemann, 1972).

28 Buckley, W. F. *Society: A Complex Adaptive System. Essays in Social Theory*, 1998).

29 Blau, P. M. *Exchange and power in social life*. (J. Wiley, 1964).

30 Schwandt, D. R. in *Complexity Leadership, Part 1: Conceptual Foundations* (eds Mary Uhl-Bein & Russ Marion) (IAP Publications, 2008).

31 Giddens, A. *The Constitution of Society*. (University of Califronia Press, 1984).

32 Ashforth, B. E. & Johnson, S. in *Social Identity Processes in Organizational Contexts* (eds M.A. Hogg & D.J. Terry) 31-48 (Psychology Press, 2001).

33 Lotrecchiano, G. R. Complexity leadership in transdisciplinary (TD) learning environments: A knowledge feedback loop. *International Journal of Transdisciplinary Research* **5**, 29-63 (2010).

34 Lotrecchiano, G. R. A dynamical approach toward understanding mechanisms of team science: Change, kinship, tension, and heritage in a transdisciplinary team. *Clinical and Translational Science* **in press** (2013).

35 Klein, J. T. Evaluation of interdiciplinary and transdiciplinary research. *American Journal of Preventative Medicine* **35**, S116-123 (2008).

36 Mead, G. H. Suggestions toward a theory of the philosophical disciplines. *Philosophical Review* **9**, 1-17 (1900).

37 Simmel, G. *Conflict*. (Free Press, 1955).

38 Buckley, W. F. *Sociology and modern systems theory*. (Prentice-Hall, 1967).

39 Scott, B. Cybernetics and the social sciences. *Systems Research and Behavioral Science* **18**, 411-420 (2001).

40 Quinn, R. E. *Beyond Rational Management: Mastering The Paradoxes and Competing Demands of High Performance*. (Jossey-Boss, 1988).

41 Harrison, M. I. & Shirom, A. *Organizational Diagnosis and Assessment: Bridging Theory and Practice*. (Sage 1999).

42 Creswell, J. W. *Qualitative Inquiry and Research Design*. (SAGE 2007).

43 Maxwell, J. A. *Qualitative Research Design*. (SAGE Publications, 2005).

44 Bandura, A. *Social Learning Theory*. (Prentice Hall, 1977).

45 Klein, J. T. The Discourse of Interdisciplinarity. *Liberal Education* **84**, 4 (1998).

46 Klein, J. T. *Interdisciplinarity : History, Theory, and Practice*. (Wayne State University Press, 1990).

47 Stones, R. *Structuration Theory*. (MacMillan, 2005).

48 Hannah, S. T. & Lester, P. B. A multilevel approach to building and leading learning organizations. *The Leadership Quarterly* **20**, 34-48 (2009).

49 Fine, H., S. . Transdisciplinarity: trying to cross boundaries. *Tamara Journal of Critical Organisation Inquiry* **6**, 16 (2007).

50 Borner, K. *et al.* A multi-level systems perspective for the science or team science. *Science Translational Medicine* **2**, 1-5 (2010).